# Study of the fundamental physical principles in atmospheric modeling based on identification of atmosphere - climate control factors

## Version 3.0 (July 2010)

## PART 1

## BROMINE EXPLOSION AT THE POLAR ARCTIC SUNRISE

## TABLE OF CONTENTS

# ABSTRACT
**Index: Gaia paradigm and biogeochemical cycling, control factors of the development of the biotic Earth, association of bromine and oxygen atmospheric chemistry, qualitative understanding of the Arctic bromine explosion, bromine pollution of the Arctic sea ice/ snow pack and atmospheric boundary layer, empirical expression of the bromine influx into atmospheric boundary layer, elaboration of ozone-oxygen transformation**

In physical-chemical domain, from the planetary to molecular scales, earth processes are the biogeochemical processes. Earth natural phenomena reflect on the control factors of the development of the biotic Earth's system. For instance, global and regional atmospheric phenomena have an origin in metabolic activities of the Earth's life. Analysis of the quite known biogeochemical processes reveals that through the biogeochemical cycling of the terminal oxidants, e.g. oxygen, nitrogen etc, superficial Earth's biota is especially influential in the directing of tropospheric processes ([4]).

The primary objective of this part of a study is to provide accumulated evidence and qualitative understanding of the associated atmospheric phenomena of the Arctic bromine explosion and their role in the functioning of the planet: we introduce the empirical expression of the bromine influx into atmospheric boundary layer and calculate total amounts of the tropospheric BrO and $Br_y$ of the Arctic origin. We establish the underlying correspondence. The next parts of a study include discussion of the biotic origin of Arctic bromine explosion and overview of the basic issues of the life-enforced development of the Earth's system and the Earth's atmosphere. All over the study, phenomena of the Arctic bromine explosion is used to illustrate the proposed concepts and ideas. In the context of the Gaia paradigm, we evaluate the ozone-oxygen transformations and their elaboration upon the Earth's life factor.

# § 1 INTRODUCTION
**Index: Tropospheric ozone field, surface ozone depletion upon halogen chemistry, statistical modeling of the Artic bromine explosion, conceptualization of the earth atmosphere as a structured physical-chemical medium upon the life force**

Several critical reviews described miscellaneous aspects of halogen chemistry in troposphere. Reviewers brought the detailed reaction pathways of chlorine, iodine and bromine species, the processing of satellite data of vertical column (VC) BrO and analysis of the observed and modeled atmospheric profiles. Halogen chemistry has received the special attention in publications dedicated to Arctic ozone depletion events (ODEs) at polar sunrise. R. von Glasow and P. Crutzen confirmed that BrO + BrO catalytic reaction cycle is a main sink for ozone in springtime polar Arctic ([8]).

For a long time, springtime Arctic bromine explosion – a natural phenomenon of exponential increase in gaseous Br radicals has been remaining a puzzle for explorers. In this paper, we examine the bromine pollution of the polar Arctic sea ice/ snow pack. By looking at the bromine "spread out" in Arctic marine boundary layer (MBL) in the context of a network with rank linkage [16], we rationalize an empirical expression for bromine flux. We conducted the simulation of phenomena for the Solar Cycle 23. Obtained by computation modeling, the amounts of BrO and $Br_y$ equated the GOME satellite data on VC BrO. The computational simulations produced for year 2001 (and 1996 omitted in this chapter) also reproduced footprints of bromine explosion and depressed ozone field downwards to the mid-latitudes of the Northern

Hemisphere (NH). Model results stay in good agreement with the field measurements ([7], [18] and [28]) and support the original idea of the unique effect of the bromine chemistry on the earth atmosphere ([8], [18], [23], [24]). Obtained results - model concentrations of chemical substances, timing of BrO cloud, vertical profiles of bromine species and ozone and their evolution, are consistent with the field findings. Acquired latitudinal and longitudinal geographical extent of model ODEs suggests a correct choice of the settings.

Based on the features of the modeled bromine explosion phenomena, we open discussion on the hypothetical physical constraint, which determines the tropospheric ozone field in the Northern Hemisphere. The constraint is represented by total amounts of the tropospheric BrO and $Br_y$ of Arctic origin and by the ratio of these total amounts. We establish the underlying correspondence between constraint and empirical coefficients in expression for Arctic bromine flux. The significance of springtime ozone depletion is in final fate of interacting chemical species. For example on local scales, ozone depleted at low altitudes can contribute to the high levels of dissolved oxygen in surface Arctic waters. Reviewing literature and model results, we hypothesize that during the bromine explosion season, total amounts of BrO and $Br_y$ of Arctic origin control the NH tropospheric $O_3$ field downward to 50 N. Although additional findings and considerations are required, it has become clear that due to impact of bromine radicals on tropospheric ozone, the annual differences in total amounts of bromine species are important for the understanding of climate variability.

We suggest that demonstrated attitudinal approach to the modeling of the distributed atmospheric phenomena can be successfully implemented in the innovative atmospheric modeling. Conceptualization of the atmosphere as a structured physical-chemical information medium under control of the Earth's life is important for the inclusion of the life driving force into the atmospheric models. Produced in the paper analysis of the life force of the Arctic bromine explosion provides an example of such necessary inclusion.

## § 2 OZONE SINK IN THE ARCTIC ATMOSPHERIC BOUNDARY LAYER
**Index:  Polar Arctic at sunrise: bromine chemistry and bromine recycling, ozone sink in atmospheric boundary layer**

Atmosphere is an integral part of nature. However, the particular atmospheric phenomenon and its study usually bring to light only common relation between a few physical – dynamical parameters and local atmospheric chemistry. Since the general relations cover a small range of parameters, they are rarely transferable to the other phenomena. Present study is an attempt to provide a complimentary atmospheric perspective by relating the particular affiliated phenomena of bromine explosion, tropospheric ozone depletion and Arctic bromine ground inventories to the biogeochemical globe cycling. BrO + BrO catalytic reaction cycle is the principal halogen sink for Arctic tropospheric ozone ([8], [11]). We address phenomenological description of phenomena associated with bromine explosion and technical issues of the bromine flux modeling in the remote Arctic region. Observational data shows a rapid character of the springtime polar Arctic ODEs. ODEs begin over the sea and near the coastal line. The depletion starts at cold and very cold temperatures. Events correlate with halogen chemistry and lead to the complete ozone destruction in the boundary layer.

Inorganic bromine compounds, especially Br, BrO, and HOBr are reactive and have the ability to destroy ozone. Barrie et al. [2] proposed a catalytic cycle of reactions. Involving a self-reaction of BrO as the rate-determining step, ozone-destructing cycle is so efficient in springtime Arctic that easily destroys any local additions to tropospheric ozone.
We use the Canadian Global Environmental Model GEM ([35], Environment Canada) to simulate springtime Arctic tropospheric ozone depletion by bromine compounds. Relevant model studies and field measurements are reviewed in order to understand spatial and temporal properties of bromine explosion and tropospheric ozone depletion. A case study narrative is brought to demonstrate bromine recycling and its impact on chemistry of the atmospheric boundary layer (ABL). The basic findings on bromine chemistry mechanism made over the last 20 years have demonstrated a number of important facts and ideas:

(1) BrO is potentially an important catalyst in the troposphere; in polar Arctic conditions, at a mixing ratio of tens of parts per trillion, BrO can eliminate all local ozone within a few days ([8] - [11], [23], [24]).
(2) It has been speculated that occurring on or near the Earth's surface, with either the snow-pack and sea ice, or marine aerosols, heterogeneous processes can autocatalytically convert sea-salt bromide or other bromine compound to gas-phase molecular bromine $Br_2$ ([19], [20]). Once in the gas phase, $Br_2$ is photolyzed, producing free bromine atoms Br, which react with ozone to form BrO. After that, BrO reacts with itself, regenerating bromine atoms and thereby completing a catalytic cycle. Catalytic cycle destroys ozone while preserving the bromine radicals. Bromine recycling on aerosols prevents termination of chain reactions as HBr recycles back to the radical form.
(3) In the springtime Arctic, bromine explosion occurs as the sunlight returns. The sunlight sharply increases the rate of $Br_2$ photolysis. Extremely high concentrations of bromine and the referred depressed ozone levels have been observed in the boundary layer. Satellite observations show the geographical extent (latitude and longitude) of depletion.
(4) The Arctic MBL is about 2 km height. Due to the fast turbulent mixing, the Arctic boundary layer is unlike the free troposphere above it. Mixing provides quick gaseous exchange with the surface of the multiyear and fresh sea ice [30].

As expected, it was able to validate GEM extended bromine chemistry version by reproducing and modulating all mentioned above findings under the settings*[1,2] established for exemplary case studies.

*[1] GEM has : Vertical layering on 28 hybrid levels; The top level is about 30 km height above the surface; The variable grid with fine resolution in Arctic (100 km by 100 km) or raw uniform grid resolution; Sea salt and sulfur aerosol production and recycling (GEM - CAM model assumptions based on Sun Gong [27], K. Toyota [31]); Several number of size bins for aerosols are presented; Once in 24 hours meteorological conditions and some chemical species are upgraded; 30 minutes model time step for updating Br2 emissions flux

*[2] List of GEM species names and aliases at
*http://collaboration.cmc.ec.gc.ca/science/rpn/gem/gem-climate/*

# § 3 FRAMEWORK FOR THE SIMULATION OF THE SURFACE OZONE DEPLETION EVENTS IN THE ARCTIC ABL
**Index: GEM model setup, bromine chemistry**

Bromine modeling includes photochemistry, gas phase reactions, heterogeneous chemistry, dry and wet deposition and transport processes modeling. Introduced to GEM, bromine chemistry is close or identical to chemistry used by other researchers on different atmospheric models ([19], [37]). Responsible for ozone depletion, bromine chemistry involves reactions occurring in gas phase and bromine species recycling on sea salt and sulfur aerosols.

In analyzed exemplary case study, the reactions added to GEM chemistry are as shown:

GAS-PHASE REACTIONS

(R1)        $Br + O_3 \rightarrow BrO$

(R2)        $BrO + BrO \rightarrow Br + Br$

(R3)        $BrO + BrO \rightarrow Br_2$

(R4)        $BrO + NO \rightarrow Br + NO_2$

(R5)        $Br + HO_2 \rightarrow HBr$

(R6)        $Br + HCHO \rightarrow HBr + CO + HO_2$

(R7)        $Br + ALD_2 \rightarrow HBr + MCO_3$

(R8)        $OH + HBr \rightarrow Br$

(R9)        $BrO + HO_2 \rightarrow HOBr$

(R10)       $BrO + NO_2 + M \rightarrow BrNO_3 + M$

(R11)       $BrO + MO_2 \rightarrow HCHO + 0.75\ HOBr + 0.25\ Br + 0.25\ HO_2$

HETEROGENEOUS REACTIONS *[3]

(R12)       $HOBr \rightarrow Br_{2\_het}$

(R13)       $BrNO_3 \rightarrow HOBr\_HNO_3\_het$

PHOTOLYSIS

(R14)       $BrO + h\nu \rightarrow Br + O$

(R15)      $Br_2 + h\nu \rightarrow Br + Br$

(R16)      $HOBr + h\nu \rightarrow Br + OH$

(R17)      $BrNO_3 + h\nu \rightarrow 0.71 * (BrO + NO_2) + 0.29 * (Br + NO_3)$

*[3] [31]: Heterogeneous reactions on CAM aerosols (g= 0.1); loss rates analysis for HOBr and BrNO3.

The computational studies ([23], [24]) had shown that only aerosol bromine production could not explain Arctic ODEs. Because our knowledge of aerosols inventories is highly uncertain and because the ground production seems to be much stronger than production on aerosols, we do not include bromine production on aerosols. The study deals with bromine recycling on aerosols as strongly suggested by T. McElroy [18]. The case study assumes an extreme (maximum) area of ground emissions. In model, we extend bromine production over the Arctic sea ice coverage.

Selected for the analysis depletion events occurred in spring 2001. For March–April of 2001, we have the reliable multiyear (MY)–fresh sea ice maps (K. Toyota; R. Kwok's CMC analysis). The differences between GEM generated sea ice maps and observed sea ice maps are negligible for the spring months. The age of the sea ice, its structure, salinity, acidity and other features were left out of consideration. For the years after 2001, MY sea ice coverage is decreasing in favor of fresh ice (P. Shepson in Schiff's lecture 2007 at York U, Canada; CMC analysis). This trend explains losses in the total area of sea ice in September. As seen from space and reported by NASA, the total area of sea ice and area of BrO cloud in late winter and spring are consistent over the past 20- year period.

## § 4 BROMINE POLLUTION OVER THE ARCTIC SEA ICE / SNOW PACK AND OVER THE ARCTIC TROPOSPHERE

**Index: Concentrating of bromine in the World Ocean and in a body of marine life, meaning of springtime ODEs for the production of the marine life**

V. Vernadsky (1965) stated that living organisms and organic matter pose the major contribution to the ocean environment by sedimentogenesis, concentrating, gaseous and oxidation- reduction processes. Concentrating of bromine in the World Ocean and marine life is a well-accepted fact. Whatever source of bromine atoms is, it is placed in the Arctic ocean. ([30]). Bromine pollution of sea ice/ snow pack is a precursor of the atmospheric phenomena of bromine explosion. While the exact association of the Arctic marine life, biogenic matter and springtime bromine explosion is hard to specify, the results of atmospheric bromine explosion are obvious. The profound results of atmospheric processing of bromine influx include (i) depressed ozone field, (ii) intensification of oxygen exchange at atmosphere - ocean barrier and (ii) transport of bromine- rich air masses to mid latitudes.

Already on the microbial level of the food webs, ozone and oxygen intakes determine survival and production of the surface-dwelling Earth's type of life. Ozone depletion events in springtime Arctic are unique because of their importance for the seasonal transition of the Northern Hemisphere's surface-dwelling biota to the productive period.

## § 4.1.  BROMINE FLUX AND TROPOSPHERIC OZONE FIELD
**Index:  Phenomenology of bromine pollution of sea ice/ snow pack, halogen activation in the heterogeneous <aqueous> medium upon sunlight, empirical formula for bromine flux into the Arctic atmospheric boundary layer**

There is a discussion on possible schemes for bromine flux in recent research works ([19], [30] and [31]). Our bromine flux scheme is built on the notable conclusions of this discussion.
In this case study, we suppose that ground bromine emissions inventories are extended over the Arctic sea ice map, including the fresh and multiyear ice:
Sources of bromine production in Arctic are under investigation and their localization is one of the goals of international buoys program (IABP, P. Shepson). Ground bromine flux resembles to come from the sea ice/ snow pack.

(1) In idealization, sea ice/ snow pack is a porous matrix consisting of ice crystals in equilibrium with brine. The detailed structures and governing processes of sea ice and snow pack are different. The pore spaces of snow pack are filled with air, containing water vapor. At warm temperature, they contain some liquid water. Halogen activation proceeds via heterogeneous reactions on ice crystal surfaces toward reactive bromine species. At this moment, we assume no distinction for bromine flux from sea ice/ snow pack.
(2) Sea ice/ snow pack surface is at the bottom level of the computational model of atmosphere. Model sea ice is covered by snow of variable depth. GEM does not manage snow pack processes, and GEM does not have chemistry at the bottom level. Bottom and adjacent to bottom levels are assumed to have the same composition. Although in the surface model layer of averaged thickness 63 m, model has some interpolation in place, how well the interpolation matches dynamics between the bottom level and the adjacent bottom level is unclear.
(3) While we suppose that bromine explosion is a result of pollution from coastal line locations, density matrix of the distributed bromine flux mostly reflects on "bromine smoothing" by diffusion. In the physical Arctic environs, diffusion redistributes bromine input to atmosphere over very large area of sea ice/ snow pack. In first episodes, when pollution is not overlapping yet very large areas, the distribution of BrO in cloud indicates on increasing distance from the spreading source. Model bromine flux is exaggerated by the largest possible area of inventories.
(4) Interpolation at GEM model bottom layer does not attempt to represent "bromine smoothing" correctly. Empirical equation for bromine flux represents interference of ozone and bromine pollution of sea ice/ snow pack; it also contains empirical threshold on the surface ozone mixing ratio.

In their extremely valuable paper ([19]), B. Michalowski et al.  limited the rate of $O_3$ depletion by the mass transfer rate of HOBr to the snow pack.  Then K. Toyota proposed the dependence of surface bromine emissions from the surface $O_3$ mixing ratio.

We transfer the qualitative understanding to the empirical formula for local $Br_2$ emissions coming from GEM generated sea ice map.

The empirical dependence for bromine flux is:

$$FLUX_{BR2} = K_{EMIS\_BR2sc} * [O_3] / 50$$

As we see, the dependence contains two scaling coefficients, $K_{EMIS\_BR2sc}$ and the dimensionless complex $[O_3] / 50$, where $O_3$ and 50 are ozone mixing ratio in units ppb (vs. initially proposed by K. Toyota $[O_3] / 40$ and $K_{EMIS\_BR2sc} = 2.64E+9$).

The expression shows a linear dependence of bromine flux on the local perturbations of ozone field. Bromine flux starts with ozone pollution and stops with full ozone destruction. The dependence points out on the net effect of bromine input to the atmosphere which is a complete depletion of the <local> surface ozone. Case study shows that the relationship is reversible for the springtime ODEs of the usual duration of 36-72 hours.

While strong depletion events finish in May, the rare and weak episodes are detected till August. Occurring at warm temperatures, longer day times and melted sea ice, these ODEs are out of the scope of our analysis. The combined effects of summer marine biological activity, upwelling and mixing of Arctic waters, atmosphere–open ocean exchange create a variety of the ozone depletion patterns. Springtime ODEs patterns are unique in means of their place in the seasonal transition of NH to the biotic productive period.

## § 4.2. RATIONALE BEHIND EMPIRICAL EXPRESSION OF BR FLUX
**Index: Consideration of the physical-chemical conditions for the polar Arctic at sunrise, "snap shot" (zero dimensional) bromine emissions, approximated solution for the bromine influx into atmosphere, biotic constraint on the toxic levels of ozone**

Different values for both scaling coefficients were used in model computations. The results of model computations cannot be proved as model independent. Even the exhibited model solution confines to the pre-criteria assumptions, the physical meaning of discussed empirical dependence for bromine flux still needs to be commented. It needs to be commented why instead of the direct dependence of $Br_2$ flux on physical fields like temperature field and others, the bromine flux equation connects the interacting compounds ($Br_2$- recycling- BrO- catalytic ozone depletion etc.). The physical theory explains what empirical values would be the best approximation to the real-life solution for distributed ground emissions. In our case, the approximation is meant to represent bromine flux over the Arctic sea ice map.

Scaling the local surface ozone mixing ratio, ppb $[O_3]$ to 50 ppb of surface ozone represents the fact that chain reactions are terminated when there is no ozone to deplete. 50 ppb is a good approximation of the termination condition. Estimated as high as 50 ppb, the mean value of the surface ozone field reflects on the ozone concentrations in tropopause and necessitate the biotic constraints on the <toxic> levels of surface and near-surface ozone.

Following observations are included in phenomenological data of surface ozone field:

(1) Surface $[O_3]$ as high as 50 ppb is observed in spring Arctic. 50 ppb of surface $[O_3]$ was reproduced by GEM for April 2001 and May 1996 for coastal locations. Boundary layer is rich of ozone due to the downward injections of stratospheric ozone.

    (2) The surface ozone mixing ratio depends on the solar radiance and temperatures. In the conditions of polar Arctic (early spring darkness, cold and very cold temperatures, when surface temperatures as low as -40 C) and steady stratified lower troposphere, chemical reactions involving sunlight are much slower [30]. To simplify, we suggest conditions are not changing significantly through the 45- 30 days run centred on the month of April. In spring of 2001, observed physical conditions were around long-term averages.

    (3) In general, the mixing ratio [$O_3$] is a model dependent value. It varies with the model amounts of ozone and model configuration, but hold for the same model (checked on 2001 and 1996 run). We expect that the dimensionless complex [$O_3$] / 50 would be capable to provide model independent information. In fact, one may consider this complex as a ratio of the local surface ozone amounts to the variance of surface ozone field, which is determined by the catalytic cycles with bromine species. The local amount or the sampling is a partial realization of the model ozone distribution. The mean value is a statistical realization of the same distribution and insensitive to any joint spatial translations within the sea ice area. Both variables completely identify the model ozone field and its skewness, although they vary by value from model to model. Physically, the ratio describes the realistic non-symmetrical ozone distribution perturbed due to the ozone ability to be depleted. *[4]

At present, we are unable to imply the live-on reliance on the cold Arctic surface temperature in GEM modeling. We could but don't make distinction between MY and fresh sea ice in terms of bromine flux (in contrast to K. Toyota's research that distinguishes between MY and fresh ice bromine emissions). As mentioned above, the given case study represent the extreme case of emissions inventories, thence scaling is fixed for total sea ice map.

Estimated as high as 50 ppb, the mean value of the global surface ozone field reflects on the sensitivity to the free oxygen (ozone and oxygen) in the surface-dwelling organic life, and especially on the sensitivity of the oxygen photosynthetic unicellular organisms and their food webs. Regional ozone and oxygen concentrations in extreme Arctic environs follow the biotic limits on the paleo marine microbiota (microbial organisms) inhabit the Arctic ocean.

*[4]  See the theory of kriging and distribution estimators in [32]

## § 5 EMPIRICAL COEFFICIENTS-BASED STRATEGY
**Index: Universality of the power law distributions, complex networks with the rank linkage satisfy power law degree distribution, abstracting of "spread over" phenomena by a complex network with the rank linkage, hierarchy of complex networks, information theory algorithms**

The interpretation of the dimensionless complex [O3] / 50 is based on the physics and chemistry of bromine production and ozone loss and stated for "immediate snap shot" of a system. Such idealistic snapshot happened at "no place in a world"; therefore, it doesn't contain reference to any geographical map. In order to get the realistic bromine flux expression, we include the diffusive "smoothing" of the bromine source over geographical space in the second scaling coefficient K $_{EMIS\_BR2sc}$.

The interpretation of the physical meaning $K_{EMIS\_BR2sc}$ is drawn out from the abstract model for a complex network with the rank linkage. For the case study of realistic bromine emissions (case Cs1), $K_{EMIS\_BR2sc}$ is approximated as 1.4 E+9 (vs. value 2.64E+9 initially suggested by K. Toyota and linked to [19]) over the total sea ice map.

For the reference case (case Cs0), $K_{EMIS\_BR2sc}$ stands for 0.01 % of realistic value and equals to 1.4 E+5. We rationalize the order of magnitude (1E+9) as the scaling uniform density of distribution. The uniform density allows to represent strength of the natural bromine source and to convert $Br_2$ flux to the meaningful physical units. We regard the $K_{EMIS\_BR2sc}$ amplitude of 1.4 ≈√2 (Pythagoras' constant) as a real dimensionless parameter in the Poisson power law equation.

In our analysis, we assume that (i) empirical coefficients in the expression for the Arctic bromine flux represent the driving force of the Earth's life and (ii) their values can be found through the abstracting natural phenomena to the processing in a complex network.

"Many different constructions are subsumed under one fundamental idea", D. Hilbert.

Prompting this expression, we bring some (classical science, information theory algorithms) considerations to back the values for empirical coefficients. Empirical parameters and constraints of the natural phenomena describe the interactions in hierarchy of the complex networks through the all available temporal-spatial sizes (dimensions). In many cases, complex networks are power law networks. As well acknowledged, power law distributions capture the chosen-size effects. Universality of the power-law distributions are based on the benefits of the complex networks of the distributed agents for the information exchange and active management. Power law distributions in the complex networks with rank linkage were found for an extremely wide class of natural and social (information exchange, information entropy) phenomena.

## § 5.1. MODELING OF THE EARTH NATURAL PHENOMENA

**Index: Conceptualization of the real-world, Earth's life as a mindset of the active life agents distributed over the Earth's system, earth atmosphere is a physical-chemical information environment; role of the atmospheric processing for the functioning of the Earth's life, power law distributions in the distributed networks**

Gödel's Incompleteness Theorem: "Anything you can draw a circle around cannot explain itself without referring to something outside the circle - something you have to assume but cannot prove."

Observed natural phenomena is a function of the number of actual trials of the events. Diverse driving forces govern the interminable development of the Earth's system and development of atmosphere, but none of these absolute forces confines to one particular conceptualization. Any particular conceptualization is based on the regular patterns in the past-time outcomes, yet patterns don't rule over random chance events.

Success of atmospheric modeling depends on our understanding of the role of atmospheric processing for the functioning of the Earth's life. In a word, the conceptualization of the real-life world depends on our assertion of the earth driving forces as deterministic and/or stochastic forces.

The predominant view of natural philosophy is that the real-life world is stochastic. Despite all, due to their statistical goals and dimensional multi scales, a majority of models is deterministic. Even though in some cases (e.g. classical mechanics phenomena), deterministic models provide sufficiently reasonable solutions, the general consistency of the real-life *statistical* driving forces has never been questioned.

Earth's atmosphere-climate system is a stochastic system. Models and computational models of the global troposphere are deterministic; models imply the fixed numerical schemes of the reversible evolution equations, that would supply the unique and stable working solution (e.g. sets of solutions for DaisyWorld) for the outcomes of the several specific patterns. It is evident that the evolution equations exist only in the anthropocentric rational perception. Deterministic atmospheric models always have a number of "screwed" parameters. Deterministic weather models are not apt to provide the long-term weather forecasts, deterministic weather-and-climate models are not actually designed to deal with forward-in-time local climate changes or the global climate transitions.

The real-life atmospheric systems are the systems undergoing the irreversible state transitions. Atmospheric systems similar to the Earth's troposphere, the Earth's stratosphere, the planetary atmospheres, are open for matter and energy exchange nonlinear systems. Geological records point out on the variations in the biogeochemical cycling. Even in the present climate conditions, atmospheric parameters straightforward point out on the variations (carbon and hydrocarbon chemical substances, etc) in the biogeochemical cycling as a driving force of the Earth's development.

In deterministic modeling, Earth's sub-systems are usually reduced to the simpler form of event-driven systems working as Finite State Machines (FSMs). FSMs are machines that tend to return relatively fast to the previous <steady stable> state or transfer to the new <steady stable> state upon the regular disturbances, spatial and temporal. The stable states of atmosphere are seasons of the cosmic cycle. For instance, the Earth's troposphere, which is balanced by the diversity of natural factors ([3], [6]) that provide troposphere's equilibrium, has four distinct seasons in the middle latitudes. For high latitudes of the globe, there is common to speak about daylight and nighttime seasons and transition periods between them.

For the observation period of 30 years, during the springtime transition, associated Arctic bromine explosion phenomena have followed the regular pattern. In this part of a study, since phenomena follow regular pattern, we infer that deterministic modeling of the phenomena can provide meaningful statistics for the BrO cloud and VC BrO.

Because Earth's life is presented by the mindset of active agents distributed over the Earth's system, it is therefore appropriate to treat natural phenomena and natural systems as processing over the complex networks of the active agents. At the seasonal physical-chemical-dynamical conditions, such complex network has a stable rank linkage. Obviously, the corresponding rank linkage is an abstract expression for the structured interactions between life agents by means of the physical-chemical medium. Apparently, such rank linkage is defined by the energy and matter transformations available for the life control. The pattern rules of linkage are such that they guarantee the system's seasonal equilibrium in the interest of the Earth's life and its development.

## § 5.2. CONCEPTUALIZATION: STEP- BY- STEP STRATEGY
**Index: Implementation of the general-science modeling strategy to the associated phenomena of Arctic bromine explosion**

In terms of the NH surface-dwelling microbiota and its food webs, the polar Arctic sunrise and the following springtime is a transitional period to the NH biotic production season**.** The case study and corresponding analysis comply with the previously listed descriptions of the bromine explosion and ODEs. We adapt one of the basic conceptualization strategies of applied science. The modeling strategy contains a number of stages, which are described below:

(1) Empirical observation of natural phenomenon –> spatial and temporal evolution of the BrO cloud detected by GOME; GOME data processed in a form of VC BrO (A. Richter, K. Toyota), ground stations' measurements

(2) Intentionally simplified, stylized abstract model of phenomenon –> processes in complex network with the rank linkage

(3) Articulation of general principles ideas, rules and patterns –> empirical correlation between $Br_2$ emissions flux and local ozone mixing ratios over the sea ice map

(4) Validation of general principles on the different levels of the model, for a example, by comparing to the well explored similar abstract models if such exist –> epidemic spread out

(5) Re - statement of model design–> acceptance of empirical dependence for $Br_2$ emissions flux from the sea ice map; parametric hypothesis has to be formulated

The realistic value for coefficient $K_{EMIS\_BR2sc}$ was approximated from a few short runs for 1996. The value fits the innovative view at troposphere as a complex networking system with rank linkage.

## § 5.3. IMPLICATIONS OF AVERAGING
**Index: Choice of random link probability in the complex networks with rank linkage, the superposition of complex networks over the physical atmospheric layers, limitations of model averaging, complex network for the Arctic bromine influx into atmosphere**

Applied mathematics has studied the efficient search algorithms in graphs where nodes are placed on a 2D lattice. Each node on the lattice has the fixed number of links whose placement is inversely correlated with the lattice distance to the other nodes. The communication graphs and the biological networks are built from the nodes that "rank" other nodes by their closeness. In graphs, nodes are connected by random links. The 2D lattice is a likely analog to the atmospheric layer.

On a wide range of spatio-temporal scales, dynamical and physical processes link atmospheric cells into the atmospheric network. As we know, the Earth's atmosphere is characterized by a very definite structure. 3-D models require the division of the model atmospheric volume to the grid cells. In GEM, numerical schemes are applied to the grid cells network. Grid cells of GEM are equivalent to the nodes of a complex network. For the variety of the distributed surface processes in the atmospheric boundary layer, model implementation on 2D lattice is a natural choice.

While the power law is applied, choice of the random link or random link probability is inverse proportional to the rank. Interestingly, networks with a linkage rank produce very short paths. As a result, networks preserve clustered local structures. These clustered local structures are similar to the structure of BrO cloud detected by GOME satellite. While BrO activation happens over the large areas, temporal separability of events may be hold up to 3 days. GOME data shows high VC BrO along the coastal line in February- March episodes. Soon afterwards, it shows a dispersed BrO cloud over the Arctic basin and subarctic regions. In this context, any two adjacent vertical layers of GEM may be considered as 3D superposition of the two 2D networks, and 3D GEM is abstracted as the superposition of twenty-eight 2D networks.

It must be explained why the superposition of macroscopic "hybrid layers" (reasoned on the atmospheric pressure gradients) and not the superposition of more complex geometry is working. Taking in account a number of the intentional simplifications, we expect that the "hybrid layers" superposition likely work for Arctic ODEs because of the succeeding significant features:

(1) Bromine emissions are mediated by the sea ice/ snow pack barrier between ocean and atmosphere. The source of bromine species is located (dispersed) on the sea ice/ snow pack (the original idea of emissions from snow pack belongs to J. McConnell ([17], [29]) and T. McElroy [18]
(2) Bromine (chemistry) completely destroys the local surface ozone
(3) The lower troposphere is stably stratified

The assumptions lead to the sequential reduction of the phenomenon representation. They allow to step down from the global atmospheric model to the abstract level of complex network *[5].

Modulated by numerical Euler - Lagrange equations, GEM atmospheric processes now take a form of the network processes. Transfer to the abstract model happens without loss or addition of any significant physical feature. However, one should understand that our analysis is biased on the averaging method approximation of dispersion.

The implications of averaging for the bromine explosion are following:

(1) Averaging method may not work for local data. (e.g., observed vs. modeled surface mixing ratios, profiles and vertical columns [$O_3$], [BrO]). For instance, the best model results are expected at coastal locations.
(2) Averaging is an apparent advantage in situations when little understanding of the physical and chemical details required, and information about local interactions is not available explicitly.

(3) Averaging holds only for the described above significant features of the natural phenomenon.

*[5] In application to the variety of the natural systems, the readable explanation on the whole mathematical subject is given by Watts( [32]), Strogatz ([25]), M. Newman ([21] and [22]), J. Kleinberg ([14] and [15]) and many others.

In certain situations, bromine production on aerosols or iodine chemistry could become significant, so a new assessment of the rank linkage must be required. In order to include an impact of bromine production on aerosols or iodine chemistry on bromine explosion/ tropospheric ozone depletion, the correctional multipliers have to be introduced to the empirical expression for the bromine flux.

## § 5.4. NETWORK WITH RANK LINKAGE
**Index: Bromine explosion vs. the epidemics spread out**

For instance, the Watts' model ([32]) assumes the social network with the number of embedded hierarchies. In the model, individuals tie to one another by exponential decay law incorporated the hierarchical identifier. One of hierarchies is corresponding to geographic location. In case of atmospheric science, there is no immediate global information on winds, temperature and pressure gradients. Hence, we are left to treat Arctic ozone depletion as following the shortest distance between the source of catalyst BrO and the emerging grid cells (i.e. nodes of network) and to introduce virtual hierarchical identifier. By analogy to the numerical experiments, where simple greedy algorithm was performed on the artificial social network with and without attrition rate*[6], we explore ozone depletion by transported bromine constituents.

Let's take a small world network of N nodes, where the average shortest path scales as ln (N). Analyzing a simple greedy geographic strategy in these networks, J. Kleinberg ([14], [15]) found boundary $(\ln N)^2$ for chain lengths. The boundary exists under certain conditions: nodes are situated on the m- dimensional lattice (our 2D flat sea ice map) with connections to their 2m closest neighbors, connections between any two nodes have the probability $p \sim r^{-m}$, where r is distance between them. In 2D topology the probability p is inversely proportional to the square of distance. In the case of HP labs [12], the geometry was not quite 2D and optimum relationship relied between $1/r^2$ and $1/r$, the later means weaker decay. In [12], the explanation was given in terms of the hierarchical identifiers.

Discussion on simulations of a random network with the power law distribution and the Poisson distribution (the same number of nodes and links for comparison) for different ranges of exponent $\tau$ ($1 < \tau < 2$, $\tau \sim 2$, $2 < \tau < 4$) is given in [1]. The empirical expression for bromine flux must display a power law or specific Poisson distribution in the node degree. This way, we are able to reflect on some grid cells (nodes) with a very high degree (sources of BrO) and many with low degree. By suggesting distributed source of bromine flux over the Arctic sea ice map, we simplified the real phenomenon of bromine explosion. In reality, bromine cloud covers large territories (GOME), but the original source of bromine is located close to or at a coastal line.

*[6] Attrition rate is a fixed probability that the node will not pass an "action message" further. In our case, a role of the "action message" is played by the air parcel with (i) high concentrations of bromine family species and (ii) the strongly depleted ozone.

Growing complexity of the large number of heterogeneous links between nodes and different groups of nodes seems to lead to increase of amount of structural properties, to the more complicated abstract network. However, from theory of information networks we know that a special network with the medium number of links ($L \sim n^{1.5}$) shows the theoretical maximum complexity. There is no surprise, we expect from the atmospheric processes to demonstrate the maximum complexity.

If we decide to analog ozone destruction to the natural infectious process, the epidemic spread out may serve as a good illustration of the "bromine smoothing".

> "In all our knowledge of general principles, what actually happens is that first of all we realize some particular application of principle, and then we realize that the particularity is irrelevant, and that there is a generality which may equally truly be affirmed", B. Russell

The description of the different scenarios of epidemic spread out in the real networks and elaborated on the threshold behavior of epidemic models was brought in [16]. The introduced there network rank $p = r(v, w)^{-\tau}$ had a meaning of the inverse probability of node v to infect w at a given time; power law had been applied. It was recognized that power value determines three different scenarios (original research as retelling by J. Kleinberg):

(i) $\tau > 2$
Epidemic spreads slowly out from the source

(ii) $1 < \tau < 2$ (best fit values for different models, often from 1.4 to 1.6)
Epidemic spreads exponentially quickly out from the source

(iii) $\tau < 1$
Epidemics spreads globally rapidly and loses memory of starting point

Reviewing of time series of VC BrO detected by GOME, one is able to say that natural bromine explosion evolves according scenario with $1 < \tau < 2$. In our context, bromine is exponentially produced (and spread out) from the source. Bromine chemistry causes to the catalytic ozone depletion (infection) of ozone, which is presented at instance in model grid cells. In order to reduce complexity, we do an important assumption about omission of other gas chemistry interferences. By this, we exclusively relate local bromine and ozone amounts.

## § 5.5. POWER LAW
**Index: Theoretic maximum of complexity for the natural networks with rank linkage**

The chemical substances are components of tropospheric chemical reactor. Atmospheric chemistry uses mixing ratios (molecular, molar, by volume, or by weight) to describe relative concentrations of the atmospheric trace gases and impurities. By definition, mixing ratio is a compositional data. Earth science adapted a view that geoscience's compositions provide information about relative, not absolute, values of components. Therefore, it was proposed that every statement about a composition would be stated in terms of ratios and logratios of components. Logratios permit the usage of scale invariance and sub compositional coherence to the full composition.

The next level of abstraction of surface processes is built by analogy to the already solved problem of stability of the inverted vertical equilibrium. Problem is well explained by Perturbation Theory*[7] (transformation of Hamiltonian to Poisson series and collection of resonant terms). The Higher - Order Perturbation Theory is a fundamental science answer on the class of natural spread out processes in real atmospheric, social and biological networks with different structures and hierarchies.

To simplify, we have suggested source of bromine flux distributed over all Arctic sea ice map. It contradicts to our knowledge that Arctic bromine inventories are located close to or at coastal line. In our expression for bromine flux, "spread out" encoded in $K_{EMIS\_BR2sc}$ as perturbation over network. Usage of Poisson law with $\tau = 1.4$ ($\approx \sqrt{2}$) is justified for the spread out of the bromine pollutants over the Arctic sea ice and compliant with the previous discussion on rank linkage and hierarchies. In order to transfer from species amounts to the actual bromine flux, we introduce logarithm of dependency. We also introduce the unit density of the distributed source of bromine. Following to [2] and [19] considerations on the actual bromine concentrations in snowmelt and K. Toyota's approximated natural Arctic fluxes for $NO_2$, HCHO and ALD2 in [31], we assume the unit density of distribution is as high as 1E+9 molec/cm$^2$/s. This value corresponds to the mass transfer per unit area per unit time. The realistic value of $K_{EMIS\_BR2sc}$ is approximated as 1.4 E+9 molec/ cm$^2$ /sec*[8]. By changing magnitude and power of $K_{EMIS\_BR2sc}$, we get model solutions, which considerably differ by (a) the total bromine amounts, (b) the distributions of species, (c) shape, extent and behavior of BrO cloud. Before the other species distributions, bromine explosion affects the tropospheric ozone field.

The realistic value for coefficient $K_{EMIS\_BR2sc}$ was approximated from a few short runs for 1996. The value fits the innovative view at troposphere as a complex networking system with rank linkage.

*[7] *http://mitpress.mit.edu/SICM/*
G.J. Sussman, J. Wisdom, *Structure and Interpretation of Classical Mechanics*, MIT Press
The reader is invited to make an independent reading and to draw his own conclusions.

*[8] This assumption is backed by model results demonstrating realistic total BrO, $Br_y$ amounts and realistic vertical profiles BrO ( Figures 4,5). The assumption is hold in conditions when iodine chemistry can be ignored.
After the realistic value for coefficient $K_{EMIS\_BR2sc}$ was approximated from a few short runs for 1996 (solar minima of the Solar Cycle 23), it was used for 2001 runs (solar maxima of the Solar Cycle 23). The value fits the innovative view at troposphere as a complex networking system with rank linkage

## § 6 CRITERIA FOR ANALYSIS
**Index:  Total amounts of the Arctic ABL BrO, tropospheric BrO/ $Br_y$ fraction**

To direct chemical <energy-matter> transformations, e.g. ozone-oxygen transformations, Earth's life exploit abiotic reactivity of the physical-chemical space. ABL atmospheric processing specifically suits to the demands imposed by the surface-dwelling microbiota, and in first of all marine microbial organisms. For the high and middle latitudes of the Northern Hemisphere, due to their relevance for the ozone-oxygen transformations in the atmospheric boundary layer, total amounts of Arctic BrO and " tropospheric BrO/ $Br_y$ fraction" are the central factors for the timing and rates of the biotic < hydrocarbon> production of the surface-dwelling Earth's life.

Essence of the atmospheric phenomena is in their statistical character: structured "heritable" information of phenomena is carried by molecules, phase content, physical and dynamical statistical parameters. A very wide range of temporal and spatial scales needs to be considered. There are a large number of chemical constituents: they can appear in gaseous, liquid and solid phases. The rates of reactions diverge considerably. In the transport processes, the daytime convective mixing is distinctly different from the night time (i) limited mixing and (ii) predominant existence of vertically stratified shear flows. Spatial scales range from the micro environments to the global environment. By the actual model runs, we demonstrate that chosen spatial and temporal resolution respond well to the physical and chemical characteristics of the non linear atmospheric processing of Arctic bromine explosion.

Any modeling study suggests certain limits and constraints and follows certain conceptions. Our understanding of the "machinery" of atmospheric chemical and physical processes remains incomplete. The physical theory of turbulence in the atmospheric boundary layer is definitely incomplete. Because bromine is distributed over the sea ice by turbulent fluxes during blowing snow, model representation of bromine pollution of sea ice/ snow pack interface implements averaging techniques and include approximated scaling coefficients.

Qualitatively and by average amounts only, the model results can be verified by observational data. Satellite column BrO gives valuable information about BrO cloud cluster features and the geographical extent of the Arctic bromine. Clustering reflects properties of the underlined ground bromine distribution. It changes through the season, indicating on variability of sources and of atmospheric dynamics. Seasonal and interannual fluctuations show that clustering information is not deducible to the ground bromine distribution.
 Although it is assumed that satellite data carries realistic statistical information on spatial and temporal separability of bromine explosion episodes, it does not provide accurate realistic local information. The satellite data as any remote sensed data carries variety of uncertainties. Processed by A. Richter, GOME data of BrO cloud carries uncertainties of the tropospheric vertical column extraction from the total vertical column BrO.  Based on the recent GEM findings and limited ground surface measurements, we assume that uncertainties are in range of some 20 %- 50%. This uncertainty allows us to use total GOME BrO to validate model tropospheric BrO and $Br_y$ amounts and, if necessary, calibrate scaling coefficients in empirical expression for bromine flux.
Probably, the major problem of the differential <calculus> modeling on the rectangular grids is potential and actual violation of the mass conservation, which introduces artificial density sinks or sources. Fortunately, GEM scheme of the surface bromine emissions demonstrated continuality of the total amounts for BrO and $Br_y$ during April 2001. Realistic $Cs1$ amounts equated the BrO amounts observed by GOME. Model profiles for the chemical substances of interest well match to the available ground observations.

We propose the main and secondary criteria for analysis. As a main criterion, we require a quantitative agreement between the total model BrO amount and BrO amount observed by satellite, and the appropriate "tropospheric" BrO/ $Br_y$ fraction. Secondary criteria include the qualitative measures such as (i) reasonable extent of ozone depletion area, (ii) reasonable spatial and temporal reconstruction of bromine explosion, (iii) reasonable BrO and $O_3$ ground mix ratios and local profiles.

## § 7 PARAMETRICAL CONSTRAINT ON THE SURFACE OZONE FIELD
**Index: Statistical behavior of the NH tropospheric ozone**

For winter and spring 1994/95, 1997, the substantial amounts of tropospheric BrO were diagnosed in polar Arctic ([7], [28]). Bromine explosion spread out toward mid latitudes (northern Spain) ([23], [24]). Compared to the model reference run Cs0, the realistic case study Cs1 stated reduction in the model zonal mean of [$O_3$]. The reduction was up to 18% in widespread areas and regionally up to 40%. GEM also calculated a large area of ozone depletion, at threshold of 1% of the maximum VC BrO. Throughout the "tracking column" summation over the chosen time period, simulated BrO cloud reached to 32% of all NH area, and $Br_y$ reaches to 44% of all NH area (non- simultaneous coverage).

We speculate that it is close to what is happened in the real-life atmosphere. By this, we argue that for the seasonal springtime transition of 1996-2001, surface bromine emissions in the Arctic region were a leading disturbance of the NH tropospheric ozone field. Natural bromine pollution brought tropospheric ozone field, and accordingly, all natural pollution system of NH, to the seasonable steady stable state.

During the bromine explosion season, we observe an interaction between the two forces driving the Earth's climate – (a) solar factor (solar radiation etc.) and (b) Earth's factor (outgassing to atmosphere etc.). Bromine outgassing to atmosphere and related tropospheric ODEs are characterized by the total amounts of BrO and $Br_y$ of Arctic origin. We propose hypothesis of the parametrical constraint on the springtime NH ozone field. Constraint is the best presented by the two measurable characteristics: (i) total amount BrO and (ii) the "tropospheric BrO/ $Br_y$ partitioning". The full range of these quantities has not been investigated yet. For 2001, the estimated amounts are shown in Figure 2.

Parametric constraint can be set up to maintain relationship between bromine and oxygen species. Constraint on the springtime ozone field allows us to understand the specific relation between bromine and oxygen biogeochemical cycles for the tropospheric processing. It is also could be explained as an realization of the Earth's life driving force - to force the Earth's troposphere behave the way the surface-dwelling Earth's life wants it to behave.

We define constraint as:

> At the springtime transition to the productive season, total amount BrO of Arctic ABL origin, and "tropospheric BrO/ $Br_y$ partitioning" together determine statistical behavior of the tropospheric ozone over the NH high and mid-latitudes *[9]

*[9] At the Arctic sunrise, this constraint establishes a connection between the bromine emissions over the Arctic sea ice map and NH tropospheric ozone field. Constraint is formulated in assumption of the normal synoptic-climatic situation. As so, the variations of solar factor,

stratospheric ozone field and sea ice coverage are expected to compensate each other. Obviously, initial $O_3$ level and the stratospheric source of $O_3$ affect statistical behavior of tropospheric ozone. In this part of a study, we omit consideration of the impact of the incoming solar radiation. Interannual variability of the bromine explosion upon solar factor is described in the second part of this study named "The hypothesis of a biotic origin for the polar sunrise Arctic bromine explosion" ([8]).

## § 8 RESEARCH PERSPECTIVES

**Index: Variations of the observed total amounts BrO; research directions, modeling of biogenic ground fluxes in reference to the prevailing tropospheric perturbations**

Our analysis of the bromine explosion provides a good scheme for the mathematical expression of distributed surface emissions, which are affecting the global climate. Because combined MY and fresh of sea ice map is less influenced by the undergoing interannual tendencies then MY sea ice alone, map for the bromine flux is preserved over the time when bromine "smoothing" is in effect. Using a suggested scheme, we are able to track changes in the ozone field related to the Arctic sea ice melting, if any.

For the past 20 years, during bromine explosion seasons, the interannual variations for the observed total amounts BrO were insignificant. Any significant variations would open way for the interesting scientific and economic developments. Possible dramatic changes in the spatial distribution and strength of the natural bromine flux immediately impact constraint's factors. Tropospheric BrO amount and tropospheric BrO/ $Br_y$ fraction can be found based on the observational and model information. Upon data analysis of the interannual time series of the measurable tropospheric BrO amount and tropospheric BrO/ $Br_y$ fraction, we can (a) answer the questions how fast Arctic climate is changing and (b) to evaluate impact of these changes in the regional Arctic climate on the status of global climate.

At the normal climate conditions, surface bromine emissions in the springtime Arctic present a leading biotic disturbance in the NH tropospheric ozone field. This disturbance heavily depends on the bromine aerosol recycling at cold temperatures and low attitudes. Recognition of the importance of the Arctic bromine explosion would stimulate research on the bromine aerosol production and recycling.

Demonstrated empirical dependencies and empirical coefficients strategy can be consequently (a) reused for the other pairs of strongly interacting chemical species and (b) utilized in the innovative atmospheric box. Proposed utilization includes the ground fluxes leading to the prevailing tropospheric perturbations.

## § 9 SUMMARY

**Index: Obstacles in exploration of the earth physical-chemical reality, functioning of the Earth's life and functioning of the Earth's system**

Analysis of the quite known biogeochemical processes reveals that life is the prime driving force of planetary development. In physical-chemical domain, from the planetary to molecular scales, earth processes are the biogeochemical processes. The biggest obstacles in the exploration of the earth processes before us today are (1) a tendency toward mechanistic interpretation of the natural processes, (2) Euclidean (3+1 space) perception, (3) observational discrepancies, (4) a lack of relentlessness. As a result, here and there, natural phenomena are considered as not affiliated with control factors of the development of the biotic Earth's system.

Atmosphere is used by the Earth's life forms to support metabolism, manage energy and transfer information between the distributed life agents. For instance, global and regional atmospheric phenomena are usually described as a result of abiotic reactivity. As contrary, through the biogeochemical cycling of the terminal oxidants, superficial Earth's biota is especially influential in the directing of tropospheric processes ([4]). Underestimation of the role of ongoing biogeochemical cycling for the development of the atmospheric-climate system, climate change and climate variability over the Earth's surface leads to the distorted interpretations of natural phenomena.

The primary objective of the current part of a study is to provide accumulated evidence and qualitative understanding of the associated atmospheric phenomena of Arctic bromine explosion and their role in the functioning (i) of the Earth's life and (ii) of the biotic Earth's system.

## § 9.1. SUMMARY ON THE ASSOCIATED PHENOMENAS OF THE ARCTIC BROMINE EXPLOSION

**Index: Real-life atmosphere as a hierarchy of complex networks of maximum complexity, impact of the halogen activation on the total column ozone**

This part of a study is consistent with similar studies on global properties of the large populations. Different disciplines intensively research the information dynamics and evolution algorithms in the network world - social and computer networking ([13], [14], [15] and [34]), epidemiological [5], statistical mechanical ([21], [22] and [25]) and many others. We argue that in order to transfer the reactants over geographical 2D or 3D map, the real atmospheric system needs to acquire a linkage between its grid cells. The linkage of the model atmospheric system is reflection on the natural forces in real atmosphere, which is the (hierarchy of) complex network(s) of maximum complexity.

In short, we provide physical justification and interpretation of the empirical expression for the surface emission processes. Empirical relationship turns out to be proportional to the linkage rank. The relationship is easy to utilize in the innovative atmospheric box or 3D models. For example, it can be used to model thermally induced processes/ reactions without introducing temperature directly. Activation of the surface bromine emissions is the most possible main sink for polar total ozone column. The seasonal rates of ozone loss are associated with atmospheric dynamics and heterogeneous chemistry at low temperatures. In winter and spring, total column $O_3$ is usually high due to the high levels of stratospheric ozone, which is defined by ozone amounts imported from the tropics and nighttime conditions. Stratospheric ozone levels are reduced due to stratospheric chemistry and due to the ultimate loss of ozone to Arctic ABL and troposphere. In Arctic ABL, ozone is completely destroyed by bromine chemistry activated by sunlight. BrO cloud spreads out over NH and continues to destroy ozone at mid-latitudes. It is open question whether biogenic bromine pollutants that have been originated in surface environment, can ascent to Arctic stratosphere and destroy stratospheric ozone. Current view is that Arctic stratosphere is bromine-polluted because of volcanic eruptions and man-made bromine emissions. We only mention the fact that accelerated in polar stratospheric clouds, halogen activation and ozone destruction by halogens proceed at low temperatures from $-60^0$C to $-40^0$C (March – April monthly mean at 30hPa), which are close to the winter surface temperatures of sea ice/ snow pack.

## § 9.2. SUMMARY ON MODELING PRINCIPLES AND STRATEGY
**Index: Effect of the ongoing biogeochemical processes on the Earth's atmosphere patterns and climate change**

The contemporary viewpoint at atmosphere, reflected in the existent global atmospheric models, is far away from incorporating the fundamental science methods and studies. For example, total energy methods in classical mechanics allow us to represent the total energy as cluster expansion, while energy is a functional of density matrix. The thermodynamic studies of sufficient conditions for stable system equilibrium allow us to make physically meaningful interpretations of local interactions and the configurational entropy, as a sum of lattice site interactions, and approximate the local response to perturbations.

Present computational models are built from bottom to top. They start from the chemical reactions and physical processes of the wide range of temporal and spatial scales, and heavily rely on model input and approximations of incomplete meteorological conditions. Tune-up of actual data and existent model algorithms affects statistical properties of output variables, and thus produces model dependent results. Based on abstract schemas and not biased on concrete meteorological input, new approach provides us with the reliable tool of reconstructing different scenarios of the planetary atmosphere history.

The chapter is limited to the case of the particular constraint on the tropospheric bromine and ozone for the springtime NH. Conceptually, constraint expresses strong relationship between climate controlling factors: biogeochemical outgassing to atmosphere and solar radiation forcing. Constraint states an interdependence of the natural surface bromine emissions and the atmospheric ozone field. It shows how the ongoing biogeochemical processes are affecting Earth's climate.

ACKNOWLEDGEMENTS

I thank K. Toyota who introduced me to the internals of GEM model and shared with me his GEM aerosol modified version. I thank J. Darewych for his kind attention. I thank M.M. who involuntary inspired my interest in atmospheric modeling and I thank my family and friends bringing sense to my life.

Animated GOME data may be seen at htt*p://www.iup.uni-bremen.de/doas/bro_from_gome.htm*
GEM information and version status may be found at
*http://collaboration.cmc.ec.gc.ca/science/rpn/gem/gem-climate/*

# § 10 FIGURES
**Index: Some model and satellite data, ground based measurements**

**Figure 1** *Sample model Br2-flux, April 2001*
Coastal areas produce significant bromine fluxes. Fewer $Br_2$-fluxes come from pole where $O_3$ depleted to near ~1ppb.

**Figure 2** *Sample GOME column BrO, April 2001*
Arctic BrO cloud is traced to NH mid- latitudes

**Figure 3** *Surface $O_3$ time series, Alert, Cs0 and Cs1*
Both cases of study give reasonable time series of ozone mix ratios $[O_3]$ at Alert, compared to local measurements. We explain the coincidence by geographic coastal location of Alert (closeness to the source of Br2 emissions). Influence of bromine aerosol recycling has not been articulated yet. Differences in surface ozone mix ratios between Cs0 and Cs1 are up to 10-25 ppb for this location.

**Figure 4** *Total model BrO & $Br_y$ amounts (Cs1) vs. GOME BrO, for April 2001*
Arctic bromine is traced to mid- latitudes. Model total amounts BrO and $Br_y$ of Arctic origin (case Cs1) are consistent with observed BrO. Observed BrO maintains constant level about 3.2E27 molec (1.7E27-3.7E27). Respectively, the model BrO is about 1.0E27-1.6E27 molec, $Br_y$ is about 4.5E27-5.5E27 molec. The "tropospheric" partitioning (ratio of total BrO amount to total $Br_y$ amount [molec]/ [molec]) over 3D space) was calculated. In steady stable state, fraction varies from 22% to 25% for Cs1 and from 12% to 16% for Cs0, while ratio thrives gradually throughout time series.

**Figure 5** Time series for bromine family partitioning at Alert ground station, surface level Cs1 vs. GOME BrO column
Some partial correlations between column BrO and surface BrO & $Br_y$ are demonstrated.

**Figure 1** *Sample model Br2-flux , April 2001*

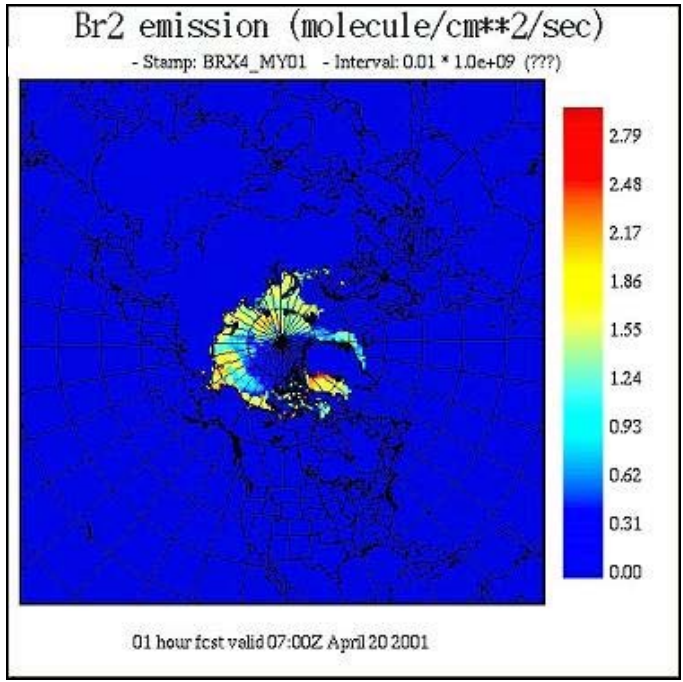

**Figure 2** *Sample GOME column BrO, April 2001( highlighted by Pisaca2)*

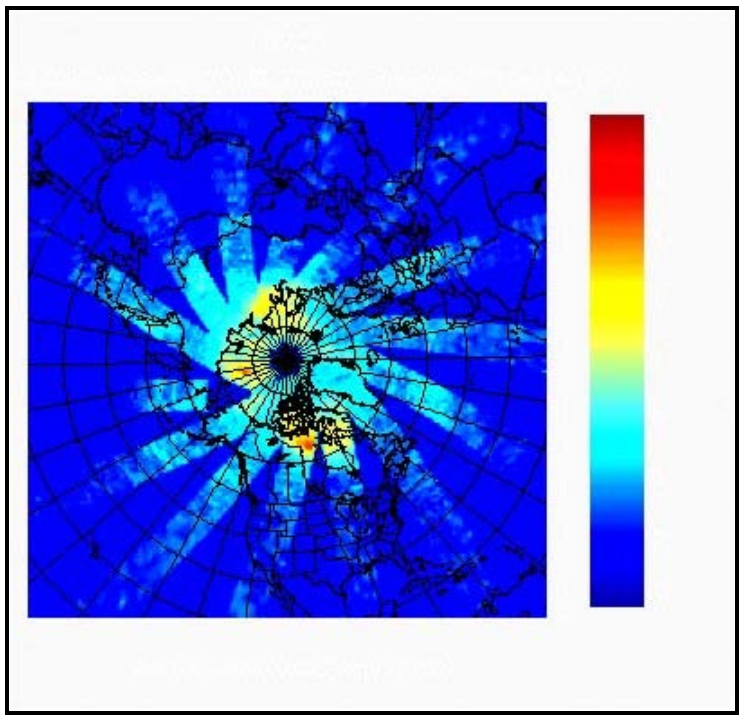

**Figure 4** *Surface O₃ time series in ppb, Alert, Cs0 and Cs1*

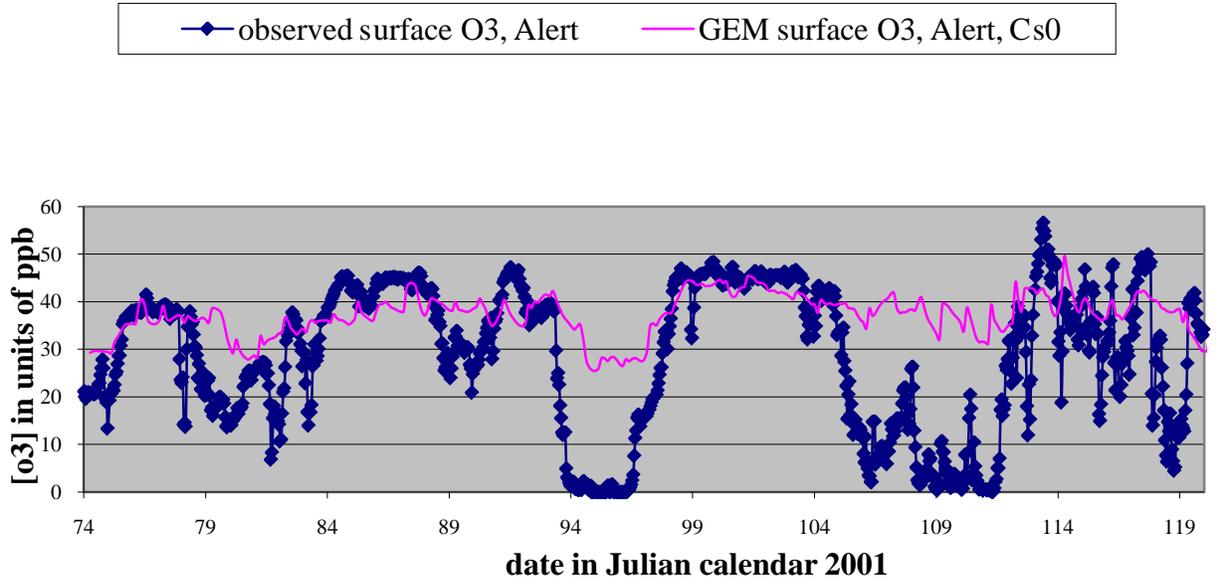

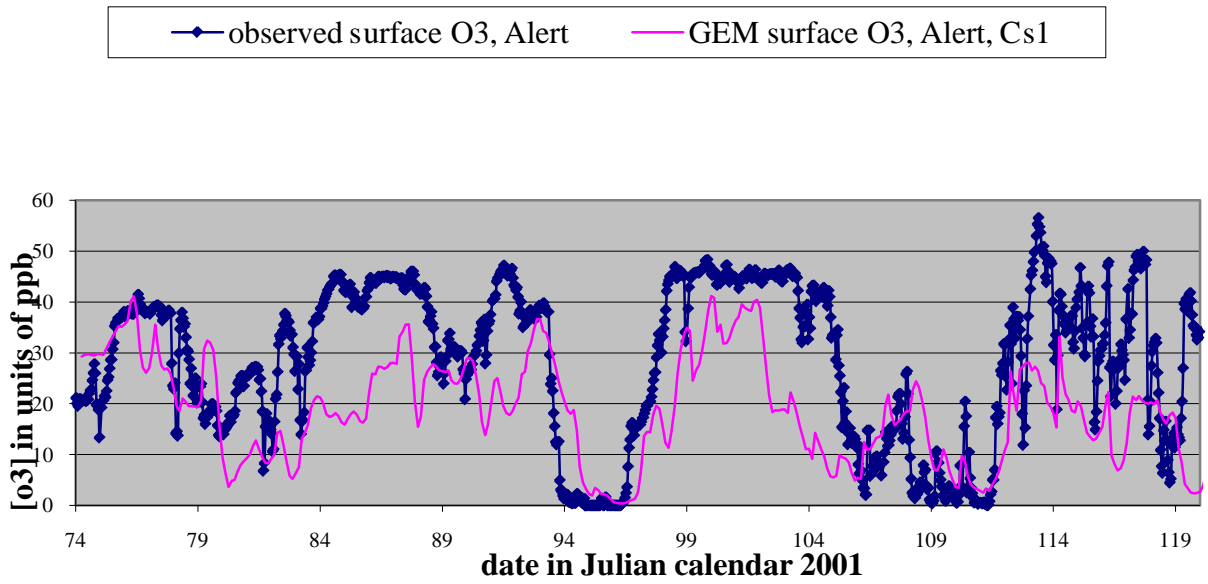

**Figure 4** *Total model BrO & $Br_y$ amounts (Cs1) vs. GOME BrO, April 2001*

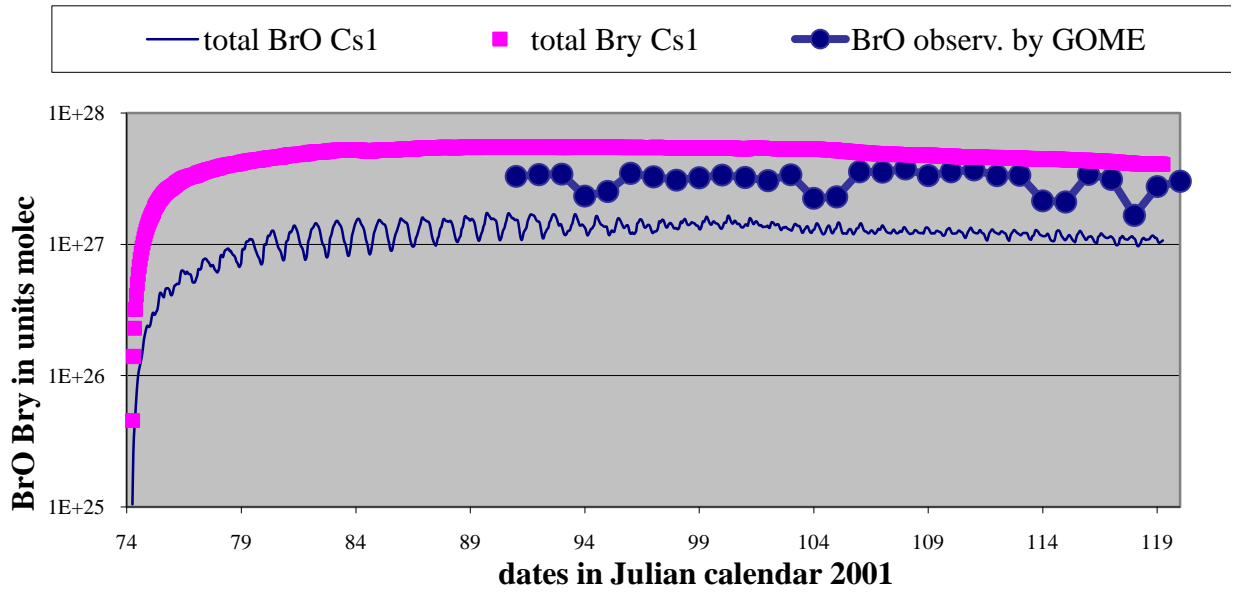

**Figure 5** *Time series for bromine family partitioning at Alert ground station, surface level, Cs1 vs. BrO column GOME*

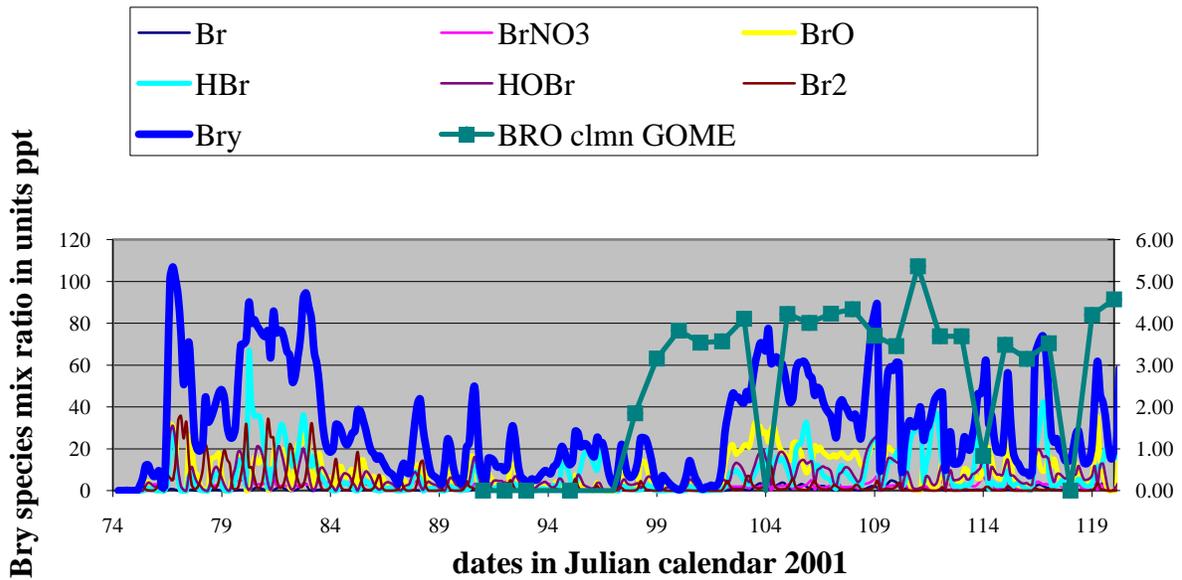

# REFERENCES

1. Adamic, L. et al., *Search in power-law networks,* Physical Review E, The American Physical Society, Vol. 64 (2001)
2. Barrie, L. A. et al. *Ozone destruction and photochemical reactions at polar sunrise in the lower Arctic atmosphere*. Nature 334, 138–141(1988).
3. Bohren, C. F., *Atmospheric Thermodynamics*. Oxford University Press (1998)
4. Crutzen, P. In: SCOPE 21, *The major biogeochemical cycles and their interactions,* SCOPE21, Wiley (1983)
5. Cointet, J. - P., Roth, C., *How Realistic Should Knowledge Diffusion Models Be.* Journal of Artificial Societies and Social Simulation, 10(3): #5 (2007)
6. Curry, J.A. &. Webster, P.J. *,Thermodynamics of Atmospheres and Oceans*. Academic Press, London (1999)
7. Friess, U. et al. *Intercomparison of measured and modelled BrO slant column amounts for the Arctic winter and spring 1994/95*, Geophys. Res. Lett., 26, 1861-1864 (1999)
8. von Glasow, R. and P. J. Crutzen, *Tropospheric Halogen Chemistry*, Holland H. D. and Turekian K. K. (eds), Treatise on Geochemistry Update1, Vol.4.02, pp 1 – 67 ( 2007)
9. von Glasow, R., Sander, R., Bott, A., Crutzen , P. J., *Modeling halogen chemistry in the marine boundary layer. 1. Cloud-free MBL*, J. Geophys. Res, 107, 4341 (2002)
10. von Glasow, R., Sander, R., Bott, A., Crutzen, P. J., *Modeling halogen chemistry in the marine boundary layer. 2. Interactions with sulfur and cloud-covered MBL*, J. Geophys. Res, 107, 4323 (2002)
11. von Glasow, R. et al., *Impact of reactive bromine chemistry in troposphere*, Atmospheric Chemistry and Physics Vol. 4, pp 2481-2497 (2004)
12. Huberman, B. A., Adamic, L. A., *Information dynamics in networked world.* arXiv:cond-mat/0308321v2, Chapter submitted to "Complex Networks", Eli Ben-Naim, Hans Frauenfelder, Zoltan Toroczkai (2003)
13. Karrer, B. et al., *Robustness of community structure in networks*, Physic Review E 77, (2008)
14. Kleinberg, J., *Navigation in a small world*, Nature, Vol. 406 (2000)
15. Kleinberg, J., *Small-world phenomena and the dynamics of information,* Advances in Neural Information Processing Systems (NIPS), Vol. 14 (2001)
*16.* Kleinberg, J. In: Keyfitz lecture, *http://www.fields.utoronto.ca/audio/07-08/keyfitz_lectures/kleinberg/*
17. McConnell, J. C. et al., *Photochemical bromine production implicated in Arctic boundary-layer ozone depletion,* Nature 355, 150–152 (1992)
18. McElroy , *http://www.nature.com/nature/journal/v397/n6717/full/397338a0.html*
19. Michalowski, B. A. et al., *A computer model study of multiphase chemistry in the Arctic boundary layer during polar sunrise*, J. Geophys. Res.*,* 105(D12), 15,131–15,146 (2001)
20. Mozurkewich, M., *Mechanisms for the release of halogens from sea salt particles by free radical reactions.* J. Geophys. Res. 100, 14199–14207 (1995)
21. Newman, M. E. J., *The structure and function of complex networks*, SIAM Review, 45(2): 167—256 (2003)
22. Newman, M. E. J., et al*., Mean- field solution of the small-world network model* (2000)

23. Sander, R. et al. *Modeling the chemistry of ozone, halogen compounds, and hydrocarbons in the arctic troposphere during spring.* Tellus, 49B, 522-532 (1997)
24. Sander, R. & Crutzen, P. J., *Model study indicating halogen activation and ozone destruction in polluted air masses transported to the sea.* J. Geophys. Res., 101D, 9121-9138 (1996)
25. Strogatz, S. H., *Exploring complex networks*, Nature 410: 268-276 (2001)
26. Seinfeld, J.H &. Pandis, S.N., *Atmospheric Chemistry and Physics*: *from Air Pollution to Global Climate Change*, Wiley, New York (1998)
27. Sun Gong, L.(Gong), L. A. Barrie, and M. Lazare (2002), *Canadian Aerosol Module (CAM): A size-segregated simulation of atmospheric aerosol processes for climate and air quality models 2. Global sea-salt aerosol and its budgets*, J. Geophys. Res., 107(D24), 4779 (2002)
28. Richter, A., F. Wittrock, M. Eisinger and J. P. Burrows: *GOME observations of tropospheric BrO in Northern Hemispheric spring and summer 1997*, Geophys. Res. Lett., No. 25, pp. 2683-2686 (1998)
29. Tang, T. & McConnell, J. C., *Autocatalytic release of bromine from Arctic snow pack during polar sunrise*. Geophys. Res. Lett. 23, 2633–2636 (1996)
30. *The tropospheric chemistry of ozone in the polar regions*, ed. Niki, H., Becker, K. New York: Springer-Verlag (1993)
31. Toyta, K., EGU 2007, presentation (2007)
32. Olea, R., *Geostatistics for Engineers and Earth Scientists*, Kluwer Academic Publishers, Holland, 328 pp (1999)
33. Vogt, R.,Crutzen ,P. J & Sander, R., *A mechanism for halogen release from sea-salt aerosol in the remote marine boundary layer.* Nature, 382, 327-330 (1996)
34. Watts, D. J. et al., *Identity and search in social networks*, Science, Vol. 296, pp. 1302–1305 (2004)
35. Plummer D., *Thesis*, http://www.yorku.ca/tropchem/thesis/appdx-a.pdf
36. Wayne, R. P. et al., *Halogen oxides: Radicals, sources and reservoirs in the laboratory and in the atmosphere*, Atmospheric Environment Vol. 29, Issue 20, Halogen oxides: Radicals, sources and reservoirs in the laboratory and in the atmosphere, pp. 2677-2881(1995)
37. Yang, X., et al., *Tropospheric bromine chemistry and its impacts on ozone: A model study*, J. Geophys. Res., 110 (2005)